\documentstyle[aps]{revtex}

\begin{document}
\title{The Global Nature of the Arrow of Time and the Bohm-Reichenbach diagram.}
\author{Mario A. Castagnino.}
\address{Instituto de Astronom\'{\i}a y F\'{\i}sica del Espacio.\\
Casilla de Correos 67, Sucursal 28,\\
1428 Buenos Aires, Argentina.}
\maketitle

\begin{abstract}
The importance of the global nature of the arrow of time is shown. Classical
Reichenbach diagram and quantum Bohm-Reichenbach diagram, for the universe
are introduced. They are used to show the increase of entropy in closed
systems, the global nature of the quantum measurement, and the relation
among the different arrows of time.
\end{abstract}

\section{ Introduction.}

This is a conceptual essay about time asymmetry with practically no
equations (these equations can be found in other contributions to this
volume or in the literature quoted in the bibliography). The main thesis is
that, even if the problem of the definition of the local arrow of time would
be completely solved (which is not yet the case since there is not an
unanimous agrement on the subject), it would not be enough to understand
time-asymmetry. In fact, the arrow of time necessarily has a global nature
and we sustain that the best structure to explain this arrow of time is a
Reichenbach global system. Let us explain these two statements:

\section{The arrow of time is global.}

Let us suppose that the arrow of time would be local. Then, it would be
possible to consider two laboratories and to define, in each one, an arrow
of time independently. Moreover let us suppose that the two laboratories are
perfectly isolated. Then we can ask ourselves: Are the two arrows of time
pointing to the same direction? Of course, this question has no answer
since, as the two laboratories are isolated, it is impossible to compare one
arrow with the other. If we would like to compare the two arrows of time an
interaction must be introduced between the two laboratories and we are
forced to consider the global system of the two laboratories and the
interaction \footnote{%
From observational evidence we know that all the laboratories in the
universe have the same arrow of time. E. g.: if not a radiastronomer would
not see the condensation of gas clouds but the opposite, an astronomers
would have found stars evolving in a direction opposite to the usual one
(Hertzsprung-Russel), or any other sign of the inequality of the local
arrows would be detected.}. We can repeat the same story if we add a third
isolated laboratory and so forth. Then {\it the} arrow of time will only be
well defined if we consider all the possible laboratories and the
interactions among each other, namely the whole universe. (See the
coincident opinion of Feynman in \cite{Feynman}).

\section{Reichenbach global system.}

The global arrow of time is best represented by a Reichenbach global system (%
\cite{Reichenbach},\cite{Davies} page 127): the system of all branching
irreversible processes within the universe, such that any process of the
system begins in an unstable state that was produced using energy coming
from another process of the global system. E. g.: the famous Gibbs ink drop
in the glass of water (initial unstable state) evolves towards a final
equilibrium state, the homogeneous mix of ink and water (final stable state)
showing that we are dealing with an irreversible process. But the ink drop
was not produced by an extremely improbable fluctuation that concentrates
the ink in the glass. It was obtained from an ink factory where, to get the
necessary energy for the factory coal (initial unstable state) was burnt in
an oven until it became ashes (final equilibrium state). {\it Furthermore
the system ''ink-water in the glass''} {\it only exists as such after the
instant when we put the ink drop into the water. }Before this instant a much
more complex system exists, that eventually contains the ink factory, the
oven, the coal burning, etc. In turn coal was not produced by a fluctuation,
quite on the contrary, it was produced using the energy coming from the sun
in geological ages. The necessary energy was provided by the light of the
sun, where H (initial unstable state) was burnt until it became He and
finally Fe (final equilibrium state). Finally H was produced using the
energy coming from the unique initial global state of the whole global
system: a cosmological initial instability. This initial unstable state can
be explained, after decoupling time, by the effect of the gravitational
field that takes the gas and radiation, in equilibrium before that time,
into a state of hot condensed clouds of matter surrounded by cold radiation,
in an expanding geometry, \cite{Aquilano}. If we want to go beyond
decoupling time we can consider the nucleosynthesis period \cite{Davies} or,
going closer to the beginning, we can consider Big-Bang quantum cosmological
models, which also have an unstable unique initial state \cite
{Hartle-Hawking}, \cite{Vilenkin}. Then through this hierarchical chain,
that begins in the cosmological instability and contains all the
irreversible processes, where each process begins where the corresponding
creation device has finished its task, the irreversible nature of the
universe and the origin of any irreversible process in it can be explained.
Therefore Gibbs ink drop only exists because there was a primordial
cosmological instability and Irreversible Statistical Mechanics can not be
explained without Irreversible Cosmology. The global system can be
symbolized as in fig. 1, which has a clear time symmetry: the branch arrow
of time (BAT), which points in opposite direction to the unique initial
cosmological instability and follows the evolution of the hierarchical chain
towards equilibrium. Reichenbach global system is clearly a realistic model
of the set of irreversible processes within the universe. Let us now see the
quantum implication of this idea.

\section{Bohm diagrams}

Let us consider a usual scattering system (with its continuous energy
spectrum as those of ref. \cite{Bohm}) and its diagram, with ingoing stable
states $a_1,$ $a_2,...$and outgoing stable states $b_1,$ $b_2,...$ and a
central black box symbolizing any interaction (fig. 2). As it is a
reversible process there is no modification of the entropy from the initial
to the final states and the eventual evolutions belong to a group. But A.
Bohm \cite{Bohm} cuts the box, at a time $t=0,$ in two pieces by a vertical
line. The l. h. s. of the this cut figure is a diagram representing the
creation of unstable states $u_1,$ $u_2,...$ from stable states $a_1,$ $%
a_2,..$ (fig. 3), with a decreasing entropy \cite{CGG}, \cite{Ordoñez}, \cite
{Edgard} and a evolution that corresponds to a creation semigroup, for $t<0$
only \cite{Bohm}, \cite{Bohm-Gadella}, \cite{Gadella}, \cite{CyLI} . The r.
h. s. of fig. 2 is a diagram representing the decaying of unstable states $%
u_1,$ $u_2,...$\footnote{%
The decaying states are obtained through the interaction of the system, with
the continuous spectrum, that plays the role of the ''environment'' \cite
{Bohm}. E. g. the system would be a H atom and the environment would be the
electromagnetic radiation that makes all energy levels unstable but the
fundamental one \cite{CyLI}. In this sense, even if the whole system is
closed (since it contains the atom and the electromagnetic radiation), it
evolves as the ''open'' system of other formalisms. As a branch system it is
therefore temporarily isolated, even if we know that a completely isolated
system does not exist. But, as explained in \cite{Davies} (page 125) it is
not the interaction with the rest of the universe (e. g. distant galaxies)
the one that produces the decaying, but the interaction with the environment
just defined.} into stable states $b_1,$ $b_2,...$with a growing entropy
that corresponds to a decaying semigroup for $t$ $>0$ (fig. 4).

The mathematical structures that correspond to the growing and decaying
processes can be essentially obtained from the quantum version of
Reichenbach idea. Let us first consider the spontaneous decaying states of
fig. 4 \footnote{%
In this period the system is quasi-isolated, and theoretically it will be
considered completely isolated, as explained in the previous footnote.}. It
is obvious that these states only exist after the creation time at $t=0$,
because before that time the system was producing growing states. Then the
probability to observe a decaying state $|\varphi (t)_{-}>$ before $t=0$ in,
e. g., any energy out-going eigenstate $|\omega _{-}>,$ namely the out-going
Lippmann-Schwinger state, is zero. The classical analogy would be to ask
what the probability is to find a particular configuration of the
distribution of the ink drop in the glass of water before the ink would be
put in the water. This probability is obviously zero since there is no ink
in the water, in complete agreement with Reichenbach idea that the
irreversible systems {\it only exist as such after the creation instant}.
Then if $t<0$ we know that $|<\omega _{-}|\varphi (t)_{-}>|^2=0,$ therefore
also $\int_{-\infty }^\infty <\omega _{-}|\varphi (t)_{-}>d\omega =0$ or $%
\int_{-\infty }^\infty <\omega _{-}|\varphi (0)_{-}>e^{-i\omega t}d\omega
=0. $ So, from the Paley-Wiener \cite{Bohm} theorem we know that $<\omega
_{-}|\varphi (0)_{-}>\in H_{+}^2$ the Hardy class from above \cite{Bohm}, 
\cite{Antoniou}, \cite{BohmPR} (see also \cite{Reisz-Simon} about the
relation of outgoing states and Hardy classes in Lax-Phillips scattering
theory). Then if we call $\Phi _{+}$ the space of states endowed with this
last property any decaying state is $|\varphi _{-}>\in \Phi _{+}$ and
decaying states can be studied considering the Gel'fand triplet $\Phi
_{+}\subset {\cal H}_{+}\subset \Phi _{+}^{\times }$ where ${\cal H}_{+}$ is
the Hilbert space outgoing states \cite{Reisz-Simon} and $\Phi _{+}^{\times }
$ is the space of antilinear functional over $\Phi _{+}.$ On the other hand,
ideal growing states \footnote{%
These states are just ideal since in the growing period the system is never
isolated but it is necessarily receiving energy from a source, i. e. from
another branch system within the global Reichenbach system. So these states
can only be considered, in the simple model of fig. 2, if we neglect the
energy source.} of fig. 3 cannot exist after $t=0$. So repeating the above
reasoning we can define the space of growing states $\Phi _{-}$ as the
states $|\varphi _{+}>$ such that $<\omega _{+}|\varphi _{+}>\in H_{-}^2,$
the Hardy class from below. Growing states can be studied using the Gel'fand
triplet $\Phi _{-}\subset {\cal H}_{-}\subset \Phi _{-}^{\times }$, where $%
{\cal H}_{-}$ is the Hilbert space incoming \cite{Reisz-Simon} states and $%
\Phi _{-}^{\times }$ is the space of antilinear functional over $\Phi _{-}.$
Using these mathematical structures the statements about semigroups and
entropy can easily be proved \cite{Edgard}. But other mathematical
structures can be used instead of the Gel'fand triplet \cite{Courbage}, \cite
{Trio}, nevertheless the semigroups and the statements about entropy always
remain the same. There are other solutions to the problem of the local arrow
of time \cite{Halliwell}, unfortunately the relation and validity of all
these solutions is not yet completely understood \cite{Ordoñez}.

\section{Bohm-Reichenbach diagram.}

From the classical Reichenbach diagram of fig. 1 and the Bohm diagrams of
the previous section we can obtain a quantum diagram for the universe, first
introduced in paper \cite{CGG}, that we will call the Bohm-Reichenbach
diagram for the universe, precisely fig. 5. It begins with the cosmological
unique primordial unstable decaying state: the r. h. s. cut box in the far
left of the figure, followed by all the scattering processes within the
universe, all connected among themselves. This global process yields a final
thermic equilibrium and therefore a growing of entropy which, defines the
thermodynamical arrow of time (TAT), that goes from the unique initial
unstable state towards equilibrium. So, from the simple inspection of the
figure, we can conclude that BAT$\equiv $TAT.

Entropy also grows in temporally closed and isolated branch systems of the
universe \footnote{%
These systems are similar to those of the previous section, so they have all
the features described in the second footnote.} as the one of the first
dotted box (A) of fig. 5, that we reproduce in fig. 6. This would be the
simplest closed branch system. It is a scattering process, such that the
outgoing states go towards equilibrium, and it also includes the source of
energy that it is used to prepare the ingoing state: the r. h. s. cut box in
the far left of the figure. The process has an overall growing of entropy
due to the decaying process of the initial cut box. Then in any realistic
(i. e. containing an irreversible process) closed subsystem of the universe
entropy always grows and the Second Law of Thermodynamics is contained, for
all closed systems, in Bohm-Reichenbach diagram.

As in the classical case, each irreversible process only exists once the
previous creating process has finished its task. Therefore the evolution of
these irreversible processes are described by semigroups beginning at the
moment of creation of each irreversible process. Therefore the universe
evolution is described by a hierarchical chain of irreversible semigroups
all based in the same Hardy class, (since all these semigroups are oriented
by the BAT, produced by the unique initial unstable states) and not by a
reversible group. Even if we have not, by now, the corresponding general
mathematical model that would prove of this statement, this scenario was
already studied in several papers \cite{Namiki}, \cite{CGL}, \cite{Lombardo}%
, \cite{CGS}, \cite{PRD} where a global space $\Phi _{+}$ is defined for the
whole universe. Moreover, it is quite logical that the global Hardy class of
the universe would endow with the same properties of analyticity all the
local spaces $\Phi _{+}$ of all the branch systems within the Reichenbach
global system.

\section{The quantum measurement process.}

Let us consider a simple example of a measurement process: Stern-Gerlach
experiment of fig 7. If we would like to consider the complete
preparation-measurement process we must consider not only the scattering
process itself, but also the accelerator ''A'' that prepares the beam, with
its source of energy, and the measurement apparatus, namely the detector and
counter ''B''. The accelerator obtains its energy from a source, where a
decaying process takes place, and in the detector a creation and a decaying
process occurs, e. g.: the particles of the deflected lower beam interact
with some atmosphere where some states are excited (creation process) and
then they decay \footnote{%
This decaying process takes place in the detector being the environment the
atmosphere within the detector. This environment is the one that transform
the closed Stern-Gerlach apparatus in an ''open'' one, in the usual
parlance. Anyhow we can consider every thing that is inside box (B) of fig.
5 as a closed system. Again the interaction with the rest of the universe
(e. g. distant galaxies) in unimportant.}. So the complete process of
preparation-measurement corresponds to the dotted box (B) of fig. 5, that we
reproduce in fig. 8. Therefore every preparation-measurement process takes
place within the Reichenbach global system, since the energy comes from a
source that can only be found in this system. Then the preparation procedure
turns out to be essentially different from the measurement one: the
preparation needs the energy that comes from the hierarchical chain, the
measurement is a decaying process, where, even if some part of the energy is
used activating the counter, the rest is degraded. Since the quantum arrow
of time (QAT) (\cite{BohmPR}, \cite{BohmPol}, \cite{BohmVol}), goes from
preparation to measurement it necessarily coincides with BAT, so QAT$\equiv $%
BAT$\equiv $TAT.

On the other hand, if Reichenbach picture is not used someone would probably
say that the difference between preparation state and measurement state is
just a technological one i. e.: the outcome of a scattering process is a
state very difficult to prepare, but anyhow, it is such that it can be
prepared with a highly refined technology. Then QAT would not be an
essential asymmetry of nature but just a technological one. This objection
disappears if we consider the preparation-measurement process within the
Reichenbach system: the difference between preparation and measurement
becomes essential: preparation needs energy coming from the primordial
instability, independently of the level of technology we use; in the
measurement process we do not need energy, which will be degraded in the
direction of the final equilibrium state of the universe. Then no confusion
is possible between quantum preparation and quantum measurement.

We also see that only in highly idealized scattering processes, with neither
preparation nor measurement (fig. 2), the evolution is described by a group.
In complete scattering processes, with the preparation and the measurement
included, the semigroup structure appears naturally.

\section{The geometrical arrow of time.}

Finally, to further emphasize the global nature of the arrow of time let us
consider the case of Classical Relativistic Cosmology. Universe is usually
considered to be a time-orientable manifold \cite{Lich}, i. e.: such that
all the null semicones can be coordinated, as in fig. 9, in order to be able
to define a global arrow of time, that points from the past semicones to the
future ones, that we will call the geometrical arrow of time (GAT). If this
would not be the case, namely if the manifold could not be oriented, we
would have non-causal loops, as O$\rightarrow $A$\rightarrow $O'$\rightarrow 
$B$\rightarrow $O of fig. 10 (where the arrows always point towards the
future, in some null cone, and nevertheless the cycle begins and ends at O)
. Then if we like to have a causal universe like ours (where, e. g.,
measurements always follow preparations) this universe must be a
time-orientable manifold and the Reichenbach system, in its relativistic
version, must be described in this manifold in such a way that GAT$\equiv $%
BAT. As a manifold can only be either orientable or non-orientable and, as
our universe is certainly a time-orientable manifold, it necessarily has a
global time-orientation, not a local one, because orientation problems are
always global (think in M\"{o}bius strip!). So GAT is global and therefore
BAT $\equiv $TAT$\equiv $QAT must also be global.

The uniqueness of the initial unstable state and the global time orientation
of the universe are, therefore, the two bases of the BAT.

We have not yet developed a mathematical structure for this heuristic model,
but a first step to introduce this structure is done in papers \cite{Pronko}%
, \cite{Alimaña} where the two semigroups are found for the time-like
directions of the null cones (one for the future and one for the past) in
Relativistic Quantum Mechanics.

\section{Conclusion.}

The morale of this essay is therefore that the arrow of time must be
considered and defined as a global object in the whole universe, most likely
using the Reichenbach global system, for the classical case, or the
Bohm-Reichenbach diagram, for the quantum case. Therefore all attempts to
define an intrinsic arrow of time based just in local reasonings are
necessarily incomplete and probably damned to failure.

\section{Acknowledgments.}

I am very grateful to the organizers of G21 for their kind invitation to
participate in this symposium. This work was partially supported by grants:
CI1$^{*}$-CT94-0004 of the European Community, PID-0150 of CONICET (National
Research Council of Argentina), EX-198 of the Buenos Aires University, and
12217/1 of Fundaci\'{o}n Antorchas and the British Council.

\section{Figures.}

Fig. 1 : The classical Reichenbach diagram.

Fig. 2: A usual scattering diagram.

Fig. 3: The creation diagram.

Fig. 4: The decaying diagram.

Fig. 5: The quantum Reichenbach diagram or Bohm-Reichenbach diagram.

Fig 6: The diagram of a close subsystem of the universe.

Fig. 7: Stern-Gerlach experiment.

Fig. 8: The Bohm-Reichenbach diagram of the Stern-Gerlach experiment.

Fig. 9: Schematic representation of an orientable manifold.

Fig. 10: Schematic representation of a non-orientable manifold with an
non-causal loop.

\end{document}